\documentclass{sigchi}

\CopyrightYear{2016}
\conferenceinfo{pre-print, submitted to ACM IUI}{}
\usepackage{balance}       % to better equalize the last page
\usepackage{graphics}      % for EPS, load graphicx instead
\usepackage[T1]{fontenc}   % for umlauts and other diaeresis
\usepackage{txfonts}
\usepackage{mathptmx}
\usepackage[pdflang={en-US},pdftex]{hyperref}
\usepackage{color}
\usepackage{booktabs}
\usepackage{textcomp}

\usepackage{microtype}        % Improved Tracking and Kerning
\usepackage{ccicons}          % Cite your images correctly!
\usepackage{subcaption}

\def\plaintitle{SIGCHI Conference Proceedings Format}

\def\emptyauthor{}
\def\plainkeywords{Authors' choice; of terms; separated; by
  semicolons; include commas, within terms only; required.}

\makeatletter
\def\url@leostyle{%
  \@ifundefined{selectfont}{
    \def\UrlFont{\sf}
  }{
    \def\UrlFont{\small\bf\ttfamily}
  }}
\makeatother
\def\pprw{8.5in}
\def\pprh{11in}

\definecolor{linkColor}{RGB}{6,125,233}
\hypersetup{%
  pdftitle={\plaintitle},
  pdfauthor={\emptyauthor},
  pdfkeywords={\plainkeywords},
  pdfdisplaydoctitle=true, % For Accessibility
  bookmarksnumbered,
  pdfstartview={FitH},
  colorlinks,
  citecolor=blue,
  filecolor=black,
  linkcolor=black,
  urlcolor=linkColor,
  breaklinks=true,
  hypertexnames=false
}

\usepackage{xspace}
\usepackage{amsmath}
\newcommand{\navTab}{Question Selection\xspace}
\newcommand{\visArea}{Documents Visualization\xspace}
\newcommand{\includedSum}{Relevant Documents Summary\xspace}
\newcommand{\selectedDetails}{Document Detail\xspace}
\newcommand{\selectedSum}{Selected Documents Summary\xspace}
\newcommand{\panel}{Selected Documents List\xspace}

\begin{document}

\title{EpistAid: Interactive Interface for Document Filtering in Evidence-based Health Care}

\numberofauthors{2}
\author{
  \alignauthor Ivania Donoso\\
    \affaddr{PUC Chile}\\
    \affaddr{Santiago, Chile}\\
    \email{indonoso@uc.cl}\\
  \alignauthor Denis Parra\\
    \affaddr{PUC Chile}\\
    \affaddr{Santiago, Chile}\\
    \email{dparras@uc.cl}\\
  % \alignauthor Author 3\\
  %   \affaddr{Affiliation}\\
  %   \affaddr{Address}\\
  %   \email{mail}\\
}

\maketitle

\begin{abstract}
Evidence-based health care (EBHC) is an important practice of medicine which attempts to provide systematic scientific evidence to answer clinical questions. In this context, \emph{Epistemonikos} ({\url{www.epistemonikos.org}})  is one of the first and most important online systems in the field, providing an interface that supports users on searching and filtering scientific articles for practicing EBHC. The system nowadays requires a large amount of expert human effort, where close to 500 physicians manually curate articles to be utilized in the platform. In order to scale up the large and continuous amount of data to keep the system updated, we introduce \emph{EpistAid}, an interactive intelligent interface which supports clinicians in the process of curating documents for \emph{Epistemonikos} within lists of papers called \textit{evidence matrices}. We introduce the characteristics, design and algorithms of our solution, as well as a prototype implementation and a case study to show how our solution addresses the information overload problem in this area.

\end{abstract}


\category{H.5.m.}{Information Interfaces and Presentation (e.g. HCI)}{Graphical User Interfaces}

%See: \url{http://www.acm.org/about/class/1998/}

\section{Introduction}

% context: EBHC potential and problems
Evidence-Based Health Care (EBHC) is a medical practice approach that emphasizes the use of research evidence to justify a medical treatment. Sackett et al. defined it as \textit{``the conscientious, explicit, and judicious use of current best evidence in making decisions about the care of individual patients''} \cite{sackett1996evidence}. %EBHC classifies information according to their ability to create high-trust recommendations.
EBHC has produced a large impact in the practice and teaching of medicine, since applying the knowledge gained from large clinical trials to patient care promotes consistency of treatment and optimal outcomes, in contrast to solely relying on habits or anecdotal cases \cite{sackett1995need}.

\begin{figure}[!t]
    \centering
    \includegraphics[width=0.85\linewidth]{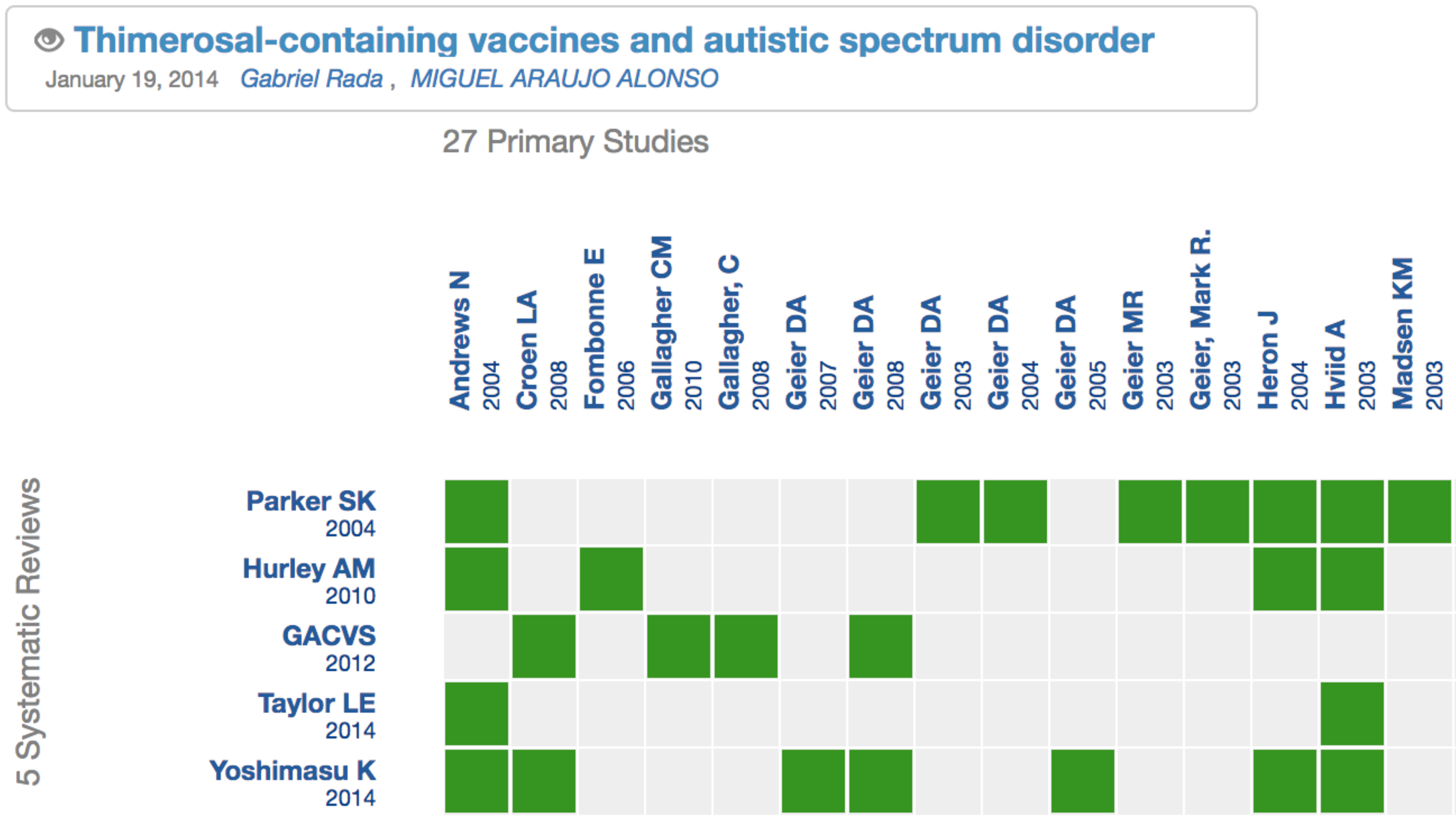}
\vspace{-2mm}
    \caption{Evidence matrix for thimerosal vaccines and autism.}
\vspace{-3mm}
    \label{fig:evidence-matrix}
\end{figure}

Despite its growing importance in health care, the process of answering a medical question is currently very expensive in time \cite{Bekhuis2014}, \cite{Miwa2014}. Clinicians pose a question, seek studies related to it, select the most relevant to the question, then perform analysis to finally obtain conclusions. This situation can be problematic because in practice, health-care related decisions must be made quickly \cite{Thomas2011}. Moreover, with the explosion of scientific knowledge being published, it is difficult for clinicians to stay updated. %on the latest best medical practices.

% \begin{figure*}[!ht]
% \centering
% \begin{subfigure}{0.5\textwidth}
%     \centering
%     \includegraphics[width=0.95\linewidth]{figures/matrix-graph-1.png}
%     \caption{Graph-based process to create an initial evidence matrix $M_0$}
%     \label{fig:matrix-graph}
% \end{subfigure}%
% \begin{subfigure}{0.5\textwidth}
%     \centering
%     \includegraphics[width=0.9\linewidth]{figures/evidence-matrix.png}
%     \caption{Example of final evidence matrix $M_f$ in \emph{Epistemonikos}}
%     \label{fig:evidence-matrix}
% \end{subfigure}
% \end{figure*}

In this context, some systems attempt to support clinicians in the process of collecting, organizing, and searching for scientific evidence such as Embase \cite{Embase}, DARE \cite{DARE}, and \textit{Epistemonikos}\cite{Rada2013}. In particular, \textit{Epistemonikos} is a collaborative database which stores research articles that provide the best evidence according to the EBHC principles \cite{Rada2013}. Since the evidence comes from scientific literature, this information is collected from specialized online sites such as PubMed %\footnote{\url{https://www.ncbi.nlm.nih.gov/pubmed/}}
 and Cochrane%\footnote{\url{http://www.cochranelibrary.com/}}
, among more than 20 other sources of scientific information.%\footnote{\url{http://www.epistemonikos.org/en/about_us/methods}}.

%Although the information in these portals is public and provides per each article data such as title, authors, abstract, year and venue of publication, the portal does not provide all the necessary information needed to answer medical questions using EBHC. First, documents need to be classified as primary studies or surveys (there are actually 4 types of surveys), and this information is not explicit in the portal. Epistemonikos has a community of physicians which manually classify articles using a crowd-sourcing procedure \citemissing{}.

% ahora las matrices de evidencia
In addition to collecting, indexing and classifying medical research articles for EBHC, \emph{Epistemonikos} developed ``Evidence Matrices'', a matrix visualization of a list of articles which provide the best evidence to answer a specific medical question, as seen in \autoref{fig:evidence-matrix}. %The creation of this matrices start with a user selecting a systematic review and the systems finds its references and, subsequently, other articles that reference it. Once all the references associated with it are found, this list is shown to the user so that \de{por norma, en lugar de he/she usa `she'} she can select the articles that are related to the question they want to answer.
Nowadays, the process of creating an evidence matrix is slow since it requires a large amount of manual and iterative effort from experts.

\begin{figure}[!t]

    \centering
    \includegraphics[width=0.9\linewidth]{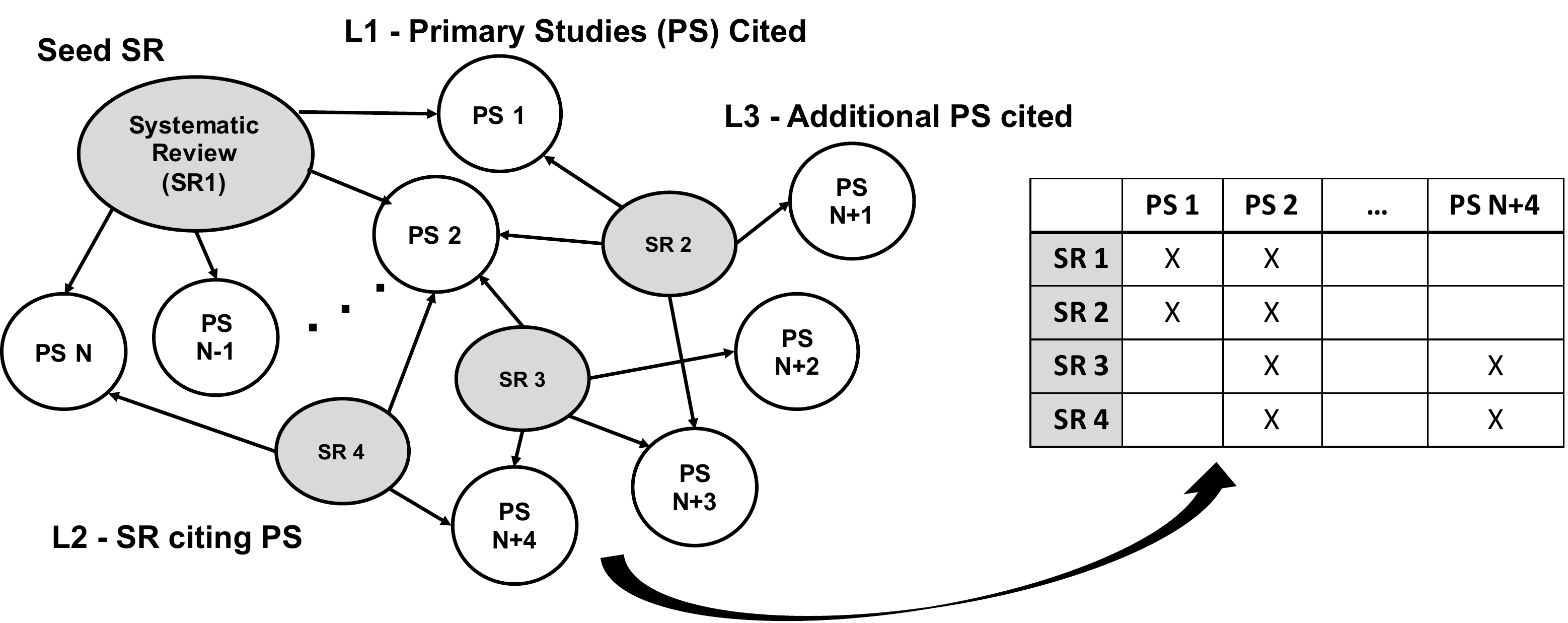}
    \caption{Graph-based process to create an initial evidence matrix $M_0$.}
    \label{fig:matrix-graph}
    \vspace{-2mm}
\end{figure}

\textbf{Contribution}. In this article, we introduce \emph{EpistAid}, a system with an intelligent user interface which support physicians in the process of creating these evidence matrices within Epistemonikos. Then, we contribute to EBHC and to the area of intelligent user interfaces by: (a) integrating dimensionality reduction and leveraging relevance feedback for assisting incremental document classification in EBHC, and (c) designing and implementing an interactive user interface which integrates the aforementioned methods to reduce the effort required to finish this important task of health care. Although previous works attempted to solve this issue automatically or with semi-supervised approaches, to the best of our knowledge this is the first work which integrates machine learning and information retrieval with an interactive user interface for EHBC. In this area, it is essential to keep a ``human in the loop'' in the process, since physicians require control and transparency while filtering documents to answer a clinical question.

\section{Building Evidence Matrices}
\label{sec:evidence-matrix}
The evidence matrix is the basic unit in \emph{Epistemonikos}. It is made of a list of papers which provide the evidence to answer a clinical question such as ``Is there a relationship between vaccines with thimerosal and autism?''. In the matrix, rows represent systematic reviews (SR) and columns are primary studies (PS) which have been cited in those SR, as seen in \autoref{fig:evidence-matrix}. PS is an umbrella term that includes any study design, qualitative or quantitative, where data is collected from individuals or groups of people. On the other side, the main objective of a SR is to synthesize primary studies.

The process of creating a \textit{final} evidence matrix $M_f$ is iterative, since involves the manual process of curating an automatically created matrix $M_0$. The method for building the initial matrix $M_0$ is shown in \autoref{fig:matrix-graph}. It starts with a user selecting a seed SR, based on a clinical question. Using Breadth-First Search over the citation graph \cite{cormen2009introduction}, all the PS cited in the seed SR are added to the matrix as columns (L1-PS). Next, other SR in the database citing the L1-PS are added as rows (L2-SR). Finally, other PS cited in the L2-SR are added as additional columns (L3-SR).

\textbf{The problem}. $M_0$ must be modified by clinicians until getting to $M_f$ by: (i) removing papers not related to the clinical question, and (ii) adding new SR and PS strongly related to the clinical question. This process can take several months, especially (ii),  since it involves manually searching and checking for papers which are not explicitly linked in the \textit{Epistemonikos}' citation database.

% \iv{No sé si updated es la mejor palabra. Mejor usamos modify?}

\textbf{Our solution}. We call our solution \emph{EpistAid}. We propose a series of methods which combined with an interactive user interface aim at reducing the time to produce $M_f$ from $M_0$. Our solution involves dimensionality reduction over the text of the articles \cite{crain2012dimensionality}, and utilizing the Rocchio algorithm for relevance feedback \cite{Manning2008IIR}.

\section{Dataset}

\emph{Epistemonikos} collects articles from 26 online sources \cite{EpisteSite}. The database contains around 370,000 documents of five types: (a) Primary Study, (b) Systematic Review, (c) Overview, (d), Structured Summary of PS, and (e) Structured Summary of SR. Among them, only (a) corresponds to specific studies, and the rest (b)-(e) are surveys of different level of detail, being (b) Systematic Review the most used. Table \ref{tab:count-type} shows the number of items per publication type in \emph{Epistemonikos}.
\begin{table}
\small
  \centering
  \begin{tabular}{l r}
    {\small\textit{Type of publication}}
    & {\small \textit{Articles in database}}\\
    \midrule
    (a) Primary Study & 261,085\\
    (b) Systematic Review & 71,597 \\
    (c) Overview & 1,068  \\
    (d) Structured Summary of PS & 1,344 \\
    (e) Structured Summary of SR & 37,735 \\
    % \bottomrule
  \end{tabular}
  \caption{Distribution of publication types in the database.}

  \label{tab:count-type}
\end{table}
Currently, there are about 2,700 public evidence matrices, but only close to 400 are in their final revised version. Physicians require 2-6 months to get from an initial version $M_0$ to a final revise $M_f$. %It takes about a day or two to filter the initial matrix $M_0$ (removing non-relevant articles), and the rest of the time is spent in finding other papers that could be related, but which were not found using the graph approach.

\begin{figure*}
  \centering
  \includegraphics[width=1.75\columnwidth]{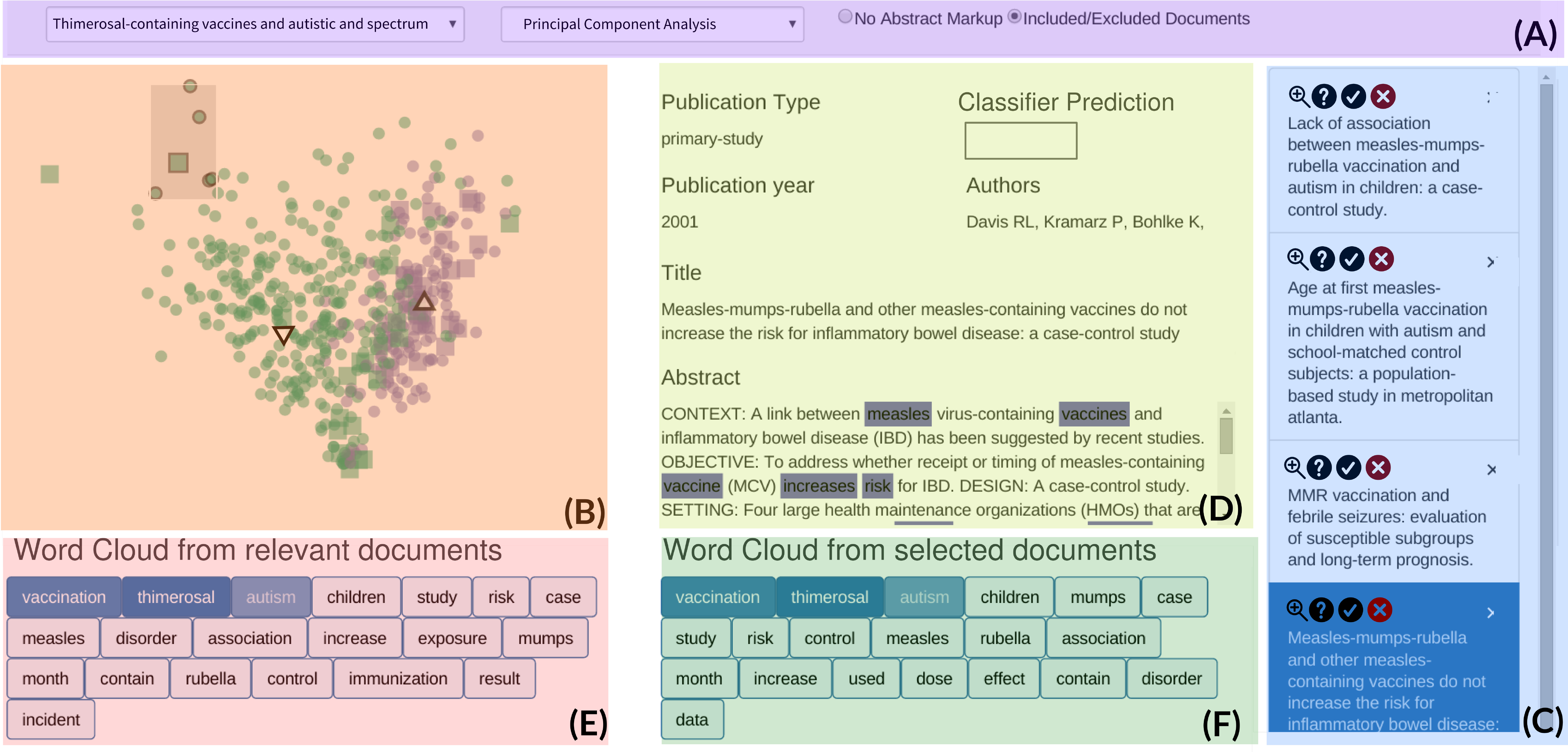}
  \caption{\emph{EpistAid} layout overview. (A) \navTab section allows selection of an evidence matrix, i.e., a document set. (B) \visArea area shows documents as figures in a 2D chart. (C) \panel section shows the selected documents list, (D) \selectedDetails shows a document's meta-data, (E) \includedSum is a wordcloud for the relevant documents and (F) \selectedSum, a wordcloud for the non-relevant documents.}
  \label{fig:main-layout}
  \vspace{-6mm}
\end{figure*}

\section{EpistAid}
Our solution encompasses a user interface and algorithms that can assist physicians during the process of filtering (removing and adding) documents related to a clinical question they want to answer. We present \emph{EpistAi}d in three parts: (i) User interface, which describes the layout and visual components, (ii) Interactions, where we justify and describe our design based on Schneidermann's visual Information-Seeking mantra \cite{Shneiderman1996}, and (iii) Algorithms, which support the intelligence behind the filtering process.

\vspace{-2mm}
\subsection{User interface}
\emph{EpistAid} user interface was developed using D3.js \cite{Bostock:2011:DDD:2068462.2068631} and Bootstrap \cite{Bootstrap}. The GUI layout, shown in \autoref{fig:main-layout}, has 6 parts and is described as follows:

\textbf{(A) \navTab.}
This navigation area lets the users choose an evidence matrix, i.e., the document set they want to see and classify. Since we represent this set of documents as a document-term matrix (DTM) using a vector space model \cite{Manning2008IIR}, we perform dimensionality reduction over this DTM to represent each document with a low-rank vector of two dimensions. Users can choose the type of dimensionality reduction (PCA, LDA, MDS \cite{engel2012survey}) they prefer in order to eventually visualize the documents in the two dimensional (2D) chart shown as \textbf{(B)}. In \autoref{fig:main-layout}, the user chose Principal Component Analysis over the documents of the evidence matrix ``Thimerosal containing vaccines and autistic spectrum disorder.''

\textbf{(B) \visArea.}
This area shows the documents as figures in a 2D chart. Its purpose is to provide an overview of the document set associated with the current evidence matrix $M_i$, and to let the user explore the content based on proximity among documents. As explained in the previous paragraph, the dimensionality reduction is chosen from the list in \textbf{(A)}.
%This area purpose is to give an overview of the current state of task and to let the user explore the content based relations among documents. The algorithm used to do the dimensionality reduction can be chosen from the list in
In the 2D chart, the primary studies (PS) are represented by circles and the systematic reviews (SR) by slightly larger squares. The color is used to discriminate the current status of the document: relevant, non-relevant or unknown. Two unfilled triangles with thick borders represent the centroids of the relevant (pointing upwards) and non-relevant documents (pointing downwards). %The current query $Q_c$ is represented by a document, the Seed Systematic Review is represented by an unfilled triangle, whose edges are thicker than the other figures.

For selecting a set of documents, the user can draw a rectangle with custom dimensions, what we call a \textit{brush}. The \textit{brush} enables the user to navigate through all the documents by sub-setting them based on their positions in the 2D projections. The selected document are displayed as a list in the right-side panel \textbf{(C)}.

\textbf{(C) \panel.}
This panel lists the documents selected with the \textit{brush} in \textit{2D Document Visualization}. When the \textit{brush} tool has not been used yet it will show the documents that have not been classified yet (unknown classification). Each item in the list shows the document title, as well as a magnifier icon that will activate the \textit{\selectedDetails} for the document, which will be displayed in panel \textbf{(D)}. Also, three icons let the user to classify the paper (as relevant, non-relevant or unknown). %, a delete button that lets the user remove the item from the selection and the document title.
%The classification button works like a carousel loop: each time the user clicks on it, the status changes accordingly. % No sé cómo explicar esto bien. La idea de este ícono es que al hacer click, se cambia al siguiente estado posible.
The documents are sorted according to their similarity to the \textit{relevance model} of the evidence matrix, a concept we explain in the next section \textbf{Algorithms}.%the current query.

% \de{cambiar nobre a DOCUMENT DETAIL.} \iv{Listo}
\textbf{(D) \selectedDetails.}
For a document selected in the \textit{\panel} \textbf{(C)} panel, this area shows its meta-data (title, abstract, type, publication year and authors) %, its current status (relevant, non-relevant, unknown)
and the classification algorithm prediction regarding its relevance. Its goal is to offer the user the option to review the documents in the same way they usually do with the current interface. %The prediction of the classifier is not shown by default to prevent a biased judgement from the author when classifying this article.
The abstract can be shown in two different ways: without markup or marking the relevant and non-relevant document keywords. In the second case the words are highlighted changing the background color with the same colors of \textit{\visArea}.

\textbf{(E) \includedSum.}
This part shows a wordcloud of important words from the current relevant documents. Its main goal is to show a quick summary of the main features (words) within this group. %It has two parts: the keywords and the publication year of the relevant documents.
%The area shows words that represent the set of documents that are already relevant.
These words can be obtained with two methods: most frequent or based on relevance \cite{Sievert2014}. % Los algoritmos los describo aquí o después? Por ahora los dejé abajo
Each word is enclosed within a button. The background color represents the relevance of the word for the user: the darker, the more important the word. The user can increase and decrease the relevance of a word to the model by clicking the right and left buttons of the mouse, respectively.

\textbf{(F) \selectedSum.}
This view has the same structure as the \textit{\includedSum} but the documents it summarizes are the ones selected with the \textit{brush} tool from \textit{\visArea}.

\subsection{Interactions}

Our interaction design is based on the visual Information-Seeking Mantra: overview first, zoom and filter, then details-on-demand \cite{Shneiderman1996}.

\textbf{Overview first.}
We implement the \textit{overview-first} functionality with the \textit{\visArea} \textbf{ (B)} where users can see a summary of the documents from the selected evidence matrix in the 2D projection resultant from a dimensionality reduction over the term-document matrix.
%Once one of the evidence matrix is selected, the system displays an overview of its documents, the \textit{\visArea} panel populated with a 2D projection of the documents in the latent space resultant from dimensionality reduction. Beneath this panel, in the \textit{\includedSum} area, there is an overview of important words from the documents that were already classified as relevant. In the \textit{\panel} area the user will see a list of documents that have not been classified yet, ordered by similarity to the Seed SR. The details of the first document of the list will be shown by default.

\textbf{Zoom and Filter: selecting documents and words.}
The \textit{brush} tool described in the previous section allows users to subset documents from the 2D \textit{\visArea}, which can be eventually analyzed in detail. When users classify documents from panel \textbf{(C)} (\textit{\panel}) and when they increase or decrease the relevance of specific words from panels \textbf{(E)} (\textit{\includedSum}) and \textbf{(F)} (\textit{\selectedSum}), they update an internal \textit{evidence matrix relevance model} which allows the system to make predictions over non-classified documents. Moreover, our system subsequently suggest documents to be reviewed by users so they can confirm the classification prediction.
%The 2D textit{\visArea} area allows the user to select a group of documents with the \textit{brush} tool already described. The documents selected can be seen in the \textit{\panel} area ordered by their similarity to the current \textit{evidence matrix relevance model}. A list of important words representing the set of documents selected will be shown below the details of the first document in the \textit{\panel} area. %These keywords will be calculated by relevance to the documents, just like in the \textit{\includedSum} area.

%The keywords that appear in the interface are buttons that can increase or decrease the importance of a word for the research question. By clicking with the left mouse button the word importance will increase, in any other case it will decrease. To show the current importance, button's opacity will change, the darker the button the more important is the word. The keywords' importance will give the classifier more information to differentiate between relevant and irrelevant. The classifier will change the predictions, the keywords that appear and the query position.

%When the user is ready to classify a document, they click the status icon until the find the class the want to assign to the document. This action will provide the classifier with more information to again change the predictions, they keywords that appear and the query position.

\textbf{Details on demand.}
By allowing the users to click in the documents on the panel (F) (\textit{\panel}), the systems provides additional details displayed in panel (D) (\textit{\selectedDetails}). In this way, it allows the user to justify her decision to classify the document.
%A document's details can be seen by clicking its magnifying glass icon in the \textit{\panel}. This will show metadata related to the document (type, publication year, authors, title, abstract), the status of the document (relevant, non-relevant, unknown) and the classifier's prediction. This last information will only be shown when the user hovers the area, to avoid bias in the classification process.
%The user can also choose to highlight the keywords presented in the abstract of the article, depending on their appearance in the relevant and non-relevant set. This feature is controlled in the \textit{\navTab}. This helps the user when a classification is difficult by showing the words that are more important for the relevant documents more easily.
% Agregar interacciones con el algoritmo
% Estas no son parte de la exploración, son más bien la interacción del usuario con el algoritmo.

\subsection{Algorithms}
% Application backend was programmed in Python using Scikit-Learn \cite{scikit-learn}, Pandas \cite{mckinney-proc-scipy-2010} and Scipy and Numpy \cite{5725236}. The documents were represented using Bag of Words. \de{las herramientas de implementacion es lo menos importante, se pone al final}  \iv{Hay que citar a alguien?}

The main aspect of our ``human-in-the-loop'' algorithmic procedure starts with modelling the evidence matrix as a \textit{query}, which is updated iteratively when users provide relevance feedback. We call this model the \textit{evidence matrix relevance model}, and it is based on Rocchio's relevance feedback algorithm \cite{Salton:1971:SRS:1102022}. We choose this model because it allows the use of feature boosting (in our case features=words) and it gives a sense of the current classification through the query.

\textbf{Relevance feedback in EpistAid}. We define a query $q_i$ as a vector of words, the weights calculated with TF-IDF as $\overrightarrow{q_i} = \{w_1, w_2, \ldots, w_n\}$
% \begin{equation}
%     \overrightarrow{q_i} = \{w_1, w_2, \ldots, w_n\}
%     \label{eq:query-model}
% \end{equation}
where $n$ is the number of words in the query $q_i$. In our system the initial query $q_0$ is made from the words in the title and abstract of the Seed SR. Afterwards, we rank the documents in the first version of the evidence matrix $M_0$ based on their cosine similarity with $q_0$, predicting the top most similar as relevant and the top most dissimilar as non-relevant. We then \textit{recommend} these documents to the users so they can confirm our predictions. Once the user provides feedback by manually classifying the recommended documents, as well as boosting or decreasing specific words from the \emph{EpistAid} interface, we update the query iteratively, such that in iteration $n$ the query is:
% \begin{multline}
%  \overrightarrow{q_n} = \alpha \overrightarrow{q_{n-1}} + \beta  \overrightarrow{q_{0}} \\
% + \gamma \frac{1}{|R|} \sum_{\overrightarrow{d_r} \in R} \overrightarrow{d_r} - \delta \frac{1}{|I|)} \sum_{\overrightarrow{d_i} \in I} \overrightarrow{d_i }
% \label{eq:query-update}
% \end{multline}
\begin{equation}
 \overrightarrow{q_n} = \alpha \overrightarrow{q_{n-1}} + \beta  \overrightarrow{q_{0}}
+ \frac{1}{|R|} \sum_{\overrightarrow{d_r} \in R} \overrightarrow{d_r}\vec{\gamma}^{T} - \frac{1}{|I|} \sum_{\overrightarrow{d_i} \in I} \overrightarrow{d_i }  \vec{\delta}^{T}
\label{eq:query-update}
\end{equation}
Where $q_{n-1}$ is the last computed query, $R$ is the set of relevant documents, $I$ is the set of non-relevant documents, $q_{0}$ is the initial query. The parameters $\alpha$, $\beta$, $\gamma$ and $\delta$ have values between $0$ and $1$. We use as parameters: $\delta = 0.1$, $\beta = 0$, $\alpha = 1$ and $\gamma = 1$ based on the values used in \cite{Salton:1971:SRS:1102022}.

\begin{figure}
    \centering
      \includegraphics[width=\columnwidth]{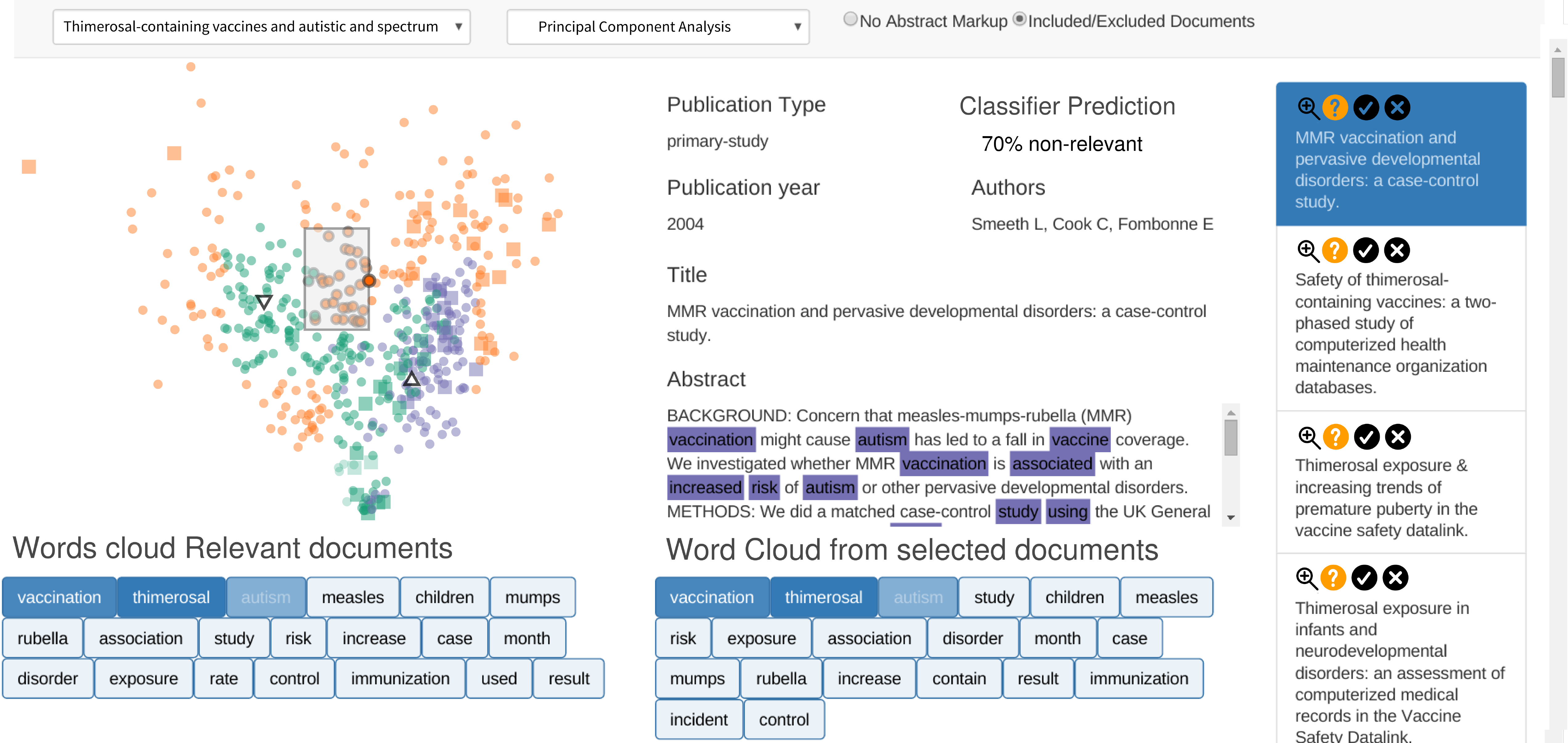}
      \caption{\textit{EpistAid} use case: modifying an evidence matrix. }
      \label{fig:intermediate}
       \vspace{-5mm}
\end{figure}

The query $q_{n}$ represents the centroid of the relevant documents, and by analogy, we can represent iteratively the centroid of the non-relevant documents $q_{n}'$. Both queries are represented visually as triangles in the 2D \textit{\visArea} panel of the \textit{EpistAid} interface, as seen in \autoref{fig:intermediate}.

\textbf{Dimensionality Reduction}
In order to display in two dimensions the term-document matrix if and evidence matrix, we chose 5 different dimensionality reduction algorithms: Principal Components Analysis (PCA), Linear Discriminant Analysis (LDA), Multidimensional Scaling (MDS) \cite{engel2012survey} and the recent t-distributed Stochastic Neighbor Embedding (t-SNE) \cite{maaten2008visualizing}. The main idea is to allow the users to visualize the status and evolution of the whole document set of an evidence matrix, as well as filtering based on their visual proximity to the relevant and non-relevant \textit{query centroids}.

% We tested them all with different document sets. The results each one gave on its own didn't gave as much information as being able to compare them. When a user make a selection of documents and changes the dimensionality reduction, the selected papers will still be highlighted. This can provide a sense of semantic relation between documents: if some documents are very close no matter the projection then we could assume that in they are very close semantically. On the opposite, when some documents change very drastically their distances with each other when the projection changes, there is a sign that the relations are not very tight.

% \subsection{Papers order}
% The \textit{\panel} was sorted according their cosine similarity to the current query.
% \iv{:O me falta alargar esta explicación}

\section{Use Case}
We present a use case of EpisteAid to filter the document set  ``Thimerosal-containing vaccines and autistic spectrum disorde''. After selecting the document set, the interface will show the documents in a two dimensional projection using PCA. The \textit{\panel} will show the all the documents, sorted by their similarity to the query. %The systematic-review that originated the document set will be classified as relevant by default. % and the current query, that corresponds to the same document will also be shown.

We will boost the words `children', `vaccination' and `autism' and classify some papers of the list. %. Figure \ref{fig:word-boost} now shows the new position of the query and the new predictions for the documents based on the new weights. Now we could classify the first 5 papers that appear on the list by clicking the icon shown in the right-top corner.
These actions will feed the system so that it can predict classes with more information.
%The user can choose to explore other documents using a brush on \textit{\visArea}. This will change the articles that appear on the \textit{\panel} area and the word cloud for selected documents. These words can be clicked to boost or decrease their weights (see figure \ref{fig:intermediate}).
The user can click the magnifying glass of a document on the \textit{\panel} area to see more details about the document. This part will also show the classifier prediction when the user hovers the area. The goal is to give its prediction, only when the user decides to prevent bias in the classification. This process continues until the classification of all the articles is completed.

\section{Related Work}

Several approaches have been proposed to reduce the workload associated with the task of document filtering for citation screening in EBHC databases. Among them we can find:
Active Learning(\cite{Figueroa2012},\cite{Miwa2014}, \cite{Rathbone2015}, \cite{Wallace2010a}, \cite{Wallace2010b}, \cite{Wallace2011}),
Automatic Classification (\cite{Bekhuis2012}, \cite{GarciaAdeva2014}, \cite{Bekhuis2014}, \cite{Mo2015}),
Document Ranking \cite{Cohen2015},
Relevance Feedback \cite{Petitti2013},
Document Priorization \cite{Cohen2009} ,
and Visualization \cite{Felizardo2012}, \cite{Felizardo2013}. The problem with the majority of these approaches is that they do not ensure 100\% recall needed to reduce the bias of the research. %With respect to these works, our work is the first one that introduces a ``human-in-the-loop'' to reduce the burden of manual document screening.
Compared to these works, we provide the first controllable and transparent information filtering system for EBHC, inspired by controllable recommender system interfaces \cite{Bostandjiev2012},\cite{Parra2015}. % Las citas hay que ordenarlas por número. No lo he hecho aún por si nos quedan citas que agregar.

%\ivde{TODO: posiscionar trabajo respecto del related work}

\section{Conclusion}
In this paper we have presented \emph{EpistAid}, a system with an interactive user interface which attempts to reduce the effort needed to curate evidence matrices in \emph{Epistemonikos}. Our upcoming work is conducting a user study with the interface and techniques described. In addition, we would like to integrate into our framework more recent techniques for relevance feedback, such as the relevance models \cite{lavrenko2001relevance}.

\section{Acknowledgments}

Authors gratefully acknowledge the grant from agency grant\#xxx--2015 (intentionally obfuscated to comply with double blind requirement)

\small

\bibliographystyle{SIGCHI-Reference-Format}
%\bibliography{sample}

\end{document}